\documentclass{ws-ijmpb}

\newcommand{\ket}[1]{\left\vert#1\right\rangle}
\newcommand{\bra}[1]{\left \langle#1\right\vert}

\newcommand{\exval}[1]{\langle{#1}\rangle}

\newcommand{\ham}{\mathcal{H}}
\DeclareMathOperator*{\Tr}{Tr}

\begin{document}

%\markboth{Authors' Names} {Instructions for Typing Manuscripts
%(Paper's Title)}

\catchline{}{}{}{}{}
%
%%%%%%%%%%%%%%%%%%%%%%%%%%%%%%%%%%%%%%%%%%%%%%%%%%%%%%%%%%%%%%%%%%%%

\title{Transport of Quantum Correlations across a spin chain}
\author{Tony J. G. Apollaro, Salvatore Lorenzo, Francesco Plastina}

\address{Dipartimento di  Fisica, Universit\`a della Calabria, 87036 Arcavacata di Rende (CS), Italy \\
INFN - Gruppo collegato di Cosenza }

\maketitle

\begin{history}
\received{Day Month Year} \revised{Day Month Year}
%\accepted{(Day Month Year)}
%\comby{(xxxxxxxxxx)}
\end{history}

\begin{abstract}
Some of the recent developments concerning the propagation of
quantum correlations across spin channels are reviewed. In
particular, we focus on the improvement of the transport
efficiency obtained by the manipulation of few energy parameters
(either end-bond strengths or local magnetic fields) near the
sending and receiving sites. We give a physically insightful
description of various such schemes and discuss the transfer of
both entanglement and of quantum discord.
\end{abstract}

\keywords{spin models; entanglement; quantum communication.}

\section{Introduction}
The problem of entanglement distribution has become of central
interest in quantum information theory and quantum communication:
quantum correlations are generated by local interactions;
therefore methods are required to transfer either the entangled
particles or their state at a distance. It has been shown
theoretically that spin chains can behave as efficient quantum
channels for short distance entanglement transfer
\cite{Bose2003,bose}, and that, for such cases, the transfer of
entanglement is strictly linked to the transport of a quantum bit
of information along the chain. As a consequence, it has become
more and more clear that the ability to manipulate the properties
of a spin chain, intended as a qubit register, can be very
important for generic communication purposes.

A simple protocol for information transmission can work as
follows: the qubit state to be transmitted is prepared at one end
of the chain, on the spin residing at the first site, and is then
propagated to the other end due to the time evolution generated by
the  spin Hamiltonian, which might be time-dependent if additional
external controls are included (see, e.g. the proposals put
forward in Ref. \cite{tdep}). Even if, for short-length chains,
the fidelity of state transfer is close to unity, it inevitably
degrades with increasing the communication distance: since the
initial state is localized, many spin excitation are typically
involved in the dynamics, causing the dispersion of the initial
information over the entire length of the chain. Thus, as this
length increases, the fidelity of state transfer substantially
reduces and various kinds of quantum correlations are generated
which spreads all over the chain length. To overcome this problem,
several schemes have been proposed to achieve perfect state
transfer, e.g. by encoding the information in low-dispersion
gaussian wave packet \cite{osbo}, by using local memories
\cite{giobu}, or by ``conclusive'' transfer, in which parallel
quantum channels are used, supplemented by a measurement at the
receiving end, allowing a high fidelity state transfer, more
robust with respect to decoherence and to non-optimal timing than
the single chain scheme \cite{bubo}.

Another possibility is to produce a refocusing of the information
at the receiving site. This can be achieved by breaking the
translational invariance of the chain,
\cite{Christandl2004,break,DiFranco2008,doronin}. In particular,
one can obtain perfect quantum transmission if the chain has
properly engineered coupling strengths and/or local fields
\cite{Christandl2004}, carefully designed in order to obtain the
so called parity matching condition \cite{shi} in the dispersion
relation of spin excitations, which guarantees a fully
constructive interference at the receiving site in the presence of
mirror inversion.

In general, indeed, the dispersion relation is non-linear; but one
can select almost-linear regions (and, therefore, get an
approximate fulfillment of the above condition) to obtain a very
high transmission fidelity by letting the state to be sent to
contain only excitations lying in this linear region
\cite{BACVV2010}.

Furthermore, it has been shown that, for such systems, one can
even relax the need for chain initialization in a reference
(fiducial) state \cite{DiFranco2008}, provided end-chain single
qubit operations can be performed. Thus, the control over the core
part of the spin medium is relaxed in favor of controllability of
the first and last element of the chain.

State transfer has been studied, both theoretically and
experimentally, with liquid-\cite{nmr1} and with solid-state NMR
\cite{nmr2} and polar molecules \cite{zhang09}, for small and
larger number of spins, respectively. In such cases, one typically
has access to many chain parameters, which can be even controlled
in time in order to achieve perfect transmission.

Another promising implementation of information transfer protocols
is obtained with trapped ions, which have been recently used to
simulate the dynamics of various spin systems \cite{lanyon}.

Different experimental implementations are expected to permit a
more restricted access to the effective Hamiltonian parameters
\cite{bu}, and therefore other methods and protocols for a high
fidelity transmission have been proposed, which rely on less
demanding control requirement\cite{gagnebin07}. One possibility is
to weaken the links between the end-spins (sender and receiver)
and the rest of the spin chain \cite{Wojcik2005,camposBO}, or to
rely on local magnetic field control \cite{plastina07}; these
schemes, however, both have the drawback of increasing the
transmission time, implying a longer exposition of the channel
itself to unwanted decoherence.

Apart from NMR-related schemes, much of the work described above
has been done with one-dimensional spin systems with short range
interactions, but chains with long-range interactions have also
been addressed \cite{longd,gualdi08}, and more general networks
have been considered \cite{jafa}, as well as the possibility to
store information \cite{storage}, rather than transmit it.

The propagation of quantum information along the chain has been
also analyzed in the presence of disorder and Anderson
localization within the spin channel\cite{disorder}, or in
presence of local environments coupled to the spins, which can
limit the distance over which quantum information can be
transmitted along the chain \cite{decoh}; moreover, environmental
correlations can play a role in such cases \cite{lorenzo11}.

\begin{figure}[bt]
\centerline{\psfig{file=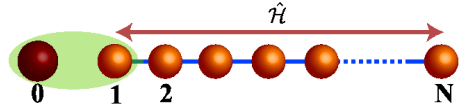,width=60mm}}
\vspace*{8pt} \caption{Sketch of a spin channel with $N$ particles
residing at the sites of a one-dimensional lattice (an open-ended
chain), whose dynamics is governed by the Hamiltonian $\hat{\cal
H}$. An external, uncoupled qubit (spin $0$) and the first spin of
the chain share some quantum correlations to be transferred
towards the other end of the chain. The pair is initially prepared
in the state $\rho^{(01)}(0)$. } \label{F.qchan}
\end{figure}

The transfer of entanglement (and, more generally, of quantum
correlations, \cite{Campbell2011}) along the chain, is strictly
related to the qubit state transfer described above. The generic
setting is sketched in Fig. \ref{F.qchan}, where the first spin of
the chain is initially prepared in an entangled state with an
external, uncoupled qubit. The aim, in this case, is to obtain, at
the end of the protocol, the greatest possible amount of
entanglement between this external qubit and the spin residing at
the last site of the chain.

In this paper we will concentrate precisely on such a process, by
reviewing various possible strategies to achieve a high quality
transmission of entanglement, based on static Hamiltonian which
include the least possible modification of the energy parameters
with respect to the case of a uniform  chain, originally treated
in Ref. \cite{Bose2003}. In fact, as mentioned above, if all of
the Hamiltonian parameters (meaning all of the coupling strengths
between the spins and the effective local magnetic fields) are
assumed to be accurately controllable, and set to specific optimal
values, perfect entanglement transfer can be achieved. The
accuracy with which the parameters need to be fixed, however,
increases with the size of the chain, and the process of
engineering the transport properties of the chain quickly becomes
very demanding. It is possible, however, to achieve a very good
transfer fidelity with low-complexity schemes in which only few
energy parameters are assumed to be controllable from outside (for
a generalization of this concept, see Ref. \cite{ghoja}, where the
general case of an on-demand transfer between any pair of selected
sites is treated). Specifically, we will focus on the cases in
which either two local fields (that is, the energy splitting of
the qubits at the sending and receiving sites) or two bond
strengths (the first and the last ones) are modified with respect
to the rest of the chain. We will see that, for a chain of any
given size, these parameters can be optimized in order to increase
the amount of transferred entanglement, still maintaining a
reasonably short transfer time.

Much of this review is dedicated to the transport of entanglement,
but we will also briefly discuss the propagation of general
quantum correlations, as quantified by the quantum discord, which,
when the system is far from its optimal transport conditions, can
be favored with respect to that of entanglement.

The remainder of this paper is organized as follows: we will first
introduce the Hamiltonian model that describes a generic
one-dimensional spin system with nearest neighbor interactions in
Sect. \ref{secgen}. Then, in Sec. \ref{xxmodel}, we will
concentrate on a specific case, the so called $XX$ model, whose
performance will be then compared with that of other anisotropic
models in Sec. \ref{animodel}. Finally, some concluding remarks
are given in Sec. \ref{conclurem}.
\section{General Hamiltonian model for the spin channel}
\label{secgen} Very different strategies have been proposed in
order to improve the transfer capability of quantum correlations
across a spin chain. In this review, we consider only static
couplings and deal with quantum channels modelled by
nearest-neighbor Heisenberg-like Hamiltonians in one dimension
\cite{nonnn}, that describe short ranged and anisotropic exchange
interaction between spins $\frac{1}{2}$ particles sitting on a
line, and subjected to effective local magnetic fields $h_i$:
\begin{equation}
\label{E.hamilton} \hat{\cal H} = \sum_{i=1}^{N-1}
J_i\left(\left(1{+}\gamma\right)\hat S^x_i \hat S^x_{i+1}
{+}\left(1{-}\gamma\right)\hat S^y_i\hat S^y_{i+1}{+}\Delta \hat
S_i^z \hat S_{i+1}^z\right) -\sum_{i=0,1} h_i \hat \sigma^z_i
\end{equation}
where $\hat S^{\alpha}_i$  ($\alpha\,{=}\,x,y,z$) are the spin
operators of the particle residing at site $i$ of an open-ended
chain, with $i\,{=}\,1,...,N$; $J_i$ is the interaction strength
between neighbor spins; $\gamma$ and $\Delta$ are the degree of
anisotropy of the coupling in the $XY${-}plane and along the
$Z$-axes, respectively, while $h$ is the magnetic field. As usual,
the eigenstates of $S^z_i$ are taken as the logical basis for
qubit $i$, $\{ \ket 0_i, \ket 1_i\}$.

This general Hamiltonian is able to describe very different
physical settings and encompasses many different features
depending on the values of the anisotropy parameters $\gamma$ and
$\Delta$. We describe various cases in the following, but mostly
concentrate on the so called $XX$-model, corresponding to
$\gamma{=}\Delta{=}0$, while leaving a brief description of the
transmission of quantum correlation in spin chains with
anisotropic couplings to the last section. There, we review the
comparison among the performances of various models, originally
performed in Ref. \cite{Bayat2011}, where it is shown that the
case of a spin chain with an isotropic coupling in the equatorial
plane and with transverse field is, in fact, the most favorable.

\section{Transport of correlations in the isotropic $XX$ model}
\label{xxmodel} Let us first consider the simplest case, known as
$XX$ model, with $\gamma{=}\Delta{=}0$.

The $XX$ Hamiltonian can be mapped, via the Jordan-Wigner
transformation, into a spinless fermion hamiltonian, up to an
irrelevant constant:
\begin{equation}\label{E.fermion}
\hat{\cal H}_{xx}{=}\sum_{i=1}^{N-1} \frac{J_i}{2} (
c^{\dagger}_{i+1}c_{i}+h.c.){-}\sum_{i=1}^{N}h_ic^{\dagger}_{i}c_{i}
\end{equation}
which is equivalent to a tight-binding hopping model in which the
fermion annihilation operator at site $i$ is related to the spin
operators as
$$c_i = \left (\prod_{j=1}^{i-1} \, 2 S^z_j  \right )\; S^-_i \, .$$
With this definition, the fermion vacuum, $\ket{\{0\}}$
corresponds to the state in which all of the spins point down. The
local single particle basis states, defined by
$\ket{i}{\equiv}c^{\dagger}_i\ket{\{0\}}{\equiv}\ket{0}^{\otimes{i-1}}\ket{1}\ket{0}^{\otimes{N-i}}$,
span the single-excitations sector of the total Hilbert space.
Restricted to this subspace and expressed in this basis, the
Hamiltonian $\hat{\cal H}$ becomes a tri-diagonal real symmetric
matrix $\mathbf{T}$ whose non-zero entries are
$t_{nm}{=}\frac{J_n}{2}\left(\delta_{n,m+1}{+}\delta_{n,m-1}\right){-}h_n\delta_{n,m}$.
The structure of the eigenvectors of $\mathbf{T}$ and the
distribution of its eigenvalues will play a central role in what
follows.

As described above, we want to consider the transfer of the
entanglement initially shared by spins $0$ and $1$ into the pair
$(0,N)$. To this end, we consider the general state
$\rho^{(0,1)}(0)$ at the initial time, and aim at obtaining an
expression for the joint state of qubits $0$ and $N$ at a
subsequent time $t$.

This latter state, $\rho^{(0N)}(t)$, can be obtained by starting
from the unitarily evolved complete density matrix of the full
spin system and then tracing out all the degrees of freedom
pertaining to spins $1, 2, \ldots, N-1$:
\begin{equation}\label{E.trace}
\rho^{(0N)}(t){=}\Tr_{\neg (0,N)}\big\{e^{-i\ham
t}\left[\rho^{(01)}(0) \otimes{\Gamma}(0)\right]e^{i\ham t}\big\}
\, ,
\end{equation}
where $\Gamma(0)$ is the initial density matrix of spins $2,
\ldots, N$.

In many cases, when investigating the transport properties over a
quantum spin  channel, one assumes that all of the spins but $0$
and $1$ are initialized in a fully polarized state; that is, spins
sitting at sites $2$ to $N$ are prepared in $\ket{0}^{\otimes
N-1}$. Because of the ${\mathcal{U}}(1)$ symmetry of the
Hamiltonian, the dynamics is then restricted to the zero- and
one-excitation subspaces and Eq.~(\ref{E.trace}) can be
straightforwardly expressed in the operator-sum
representation~\cite{Bose2003}:
\begin{equation}\label{E.krauss}
\rho^{(0N)}(t){=}\sum_{i=0,1}\left(\mathbf{1}\otimes\mathbf{M}_{i}(t)\right)
\rho^{(01)}(0)\left(\mathbf{1}\otimes\mathbf{M}_{i}(t)\right)^{\dagger},
\end{equation}
where $\mathbf{1}$ is the $2$x$2$ identity matrix acting on the
isolated qubit $0$, while $\mathbf{M}_{i}(t)$ are super-operators
describing the amplitude damping process of qubit 1:
\begin{equation*}
\mathbf{M}_{0}(t)=\left[\begin{matrix}
1 & 0  \\
0 & u_{N1}(t) \\
\end{matrix}\right]\,\,\,;\,\,\,
\mathbf{M}_{1}(t)=\left[\begin{matrix}
0 & \sqrt{1-\left|u_{N1}(t)\right|^2}  \\
0 & 0 \\
\end{matrix}\right] \, .
\end{equation*}
Here the only relevant dynamical parameter entering the Kraus
operators is the transition amplitude of the excitation from site
1 to site $N$, that is, $u_{N1}(t)\,{=}\,\bra{N} e^{-i\hat{\cal H}
t}\ket{1}$.

We consider initial input states  for the two qubits $(0,1)$
having an $X$-type form of the density matrix
\begin{equation}
\label{E.Xrho} \rho^{(01)}(0)=\left[\begin{matrix}
 \rho_{11} & 0 & 0 & \rho_{14} \\
0 & \rho_{22} & \rho_{23} & 0 \\
0 & \rho^*_{23} & \rho_{33} & 0 \\
 \rho^*_{14} & 0 & 0 & \rho_{44}
\end{matrix}\right]~~~\text{with}~\sum^4_{j=1}\rho_{jj}{=}1;
\end{equation}
then, thanks to Eq.~(\ref{E.krauss}), the $X$-type nature is
preserved and the only non-zero elements of $\rho^{0N}(t)$ are
given by
\begin{equation}
\begin{aligned}
\rho^{(0N)}_{11}(t)&\,{=}\,\rho_{11}\,{+}\,(1\,{-}\,|u_{N1}(t)|^2)\rho_{22}\,\,\,,\,\,\,\rho^{(0N)}_{22}(t)\,{=}\,|u_{N1}(t)|^2\rho_{22},\\
%&+(1-|f_{r}_{r's'}(t)|^2)(1-|f_{r}_{rs}(t)|^2)\rho_{44},\\
\rho^{(0N)}_{33}(t)&\,{=}\,\rho_{33}\,{+}\,(1\,{-}\,|u_{N1}(t)|^2)\rho_{44}\,\,\,,\,\,\,\rho^{(0N)}_{44}(t)\,{=}\,|u_{N1}(t)|^2\rho_{44},\\
%\rho^{(r0)}_{44}(t)=|f_{r}(t)|^2\rho_{44},\\
\rho^{(0N)}_{14}(t)&\,{=}\,u_{N1}(t) \rho_{14}\,\,\,,\,\,\,
\rho^{(0N)}_{23}(t)\,{=}\,u_{N1}(t) \rho_{23}.
\end{aligned}\label{e.rhooutrr'}
\end{equation}

The Concurrence $C$ between any two qubit whose density matrix can
be expressed in the $X$-form of Eq.~(\ref{E.Xrho}), takes a
particulary simple expression~\cite{AmicoOPFM2004}:
$C{=}2\max[0,\left|\rho_{23}\right|{-}\sqrt{
\rho_{11}\rho_{44}},\left|\rho_{14}\right|{-}\sqrt{
\rho_{22}\rho_{33}}]$. For the case in which the  input state
$\rho^{(01)}$ is maximally entangled, the Concurrence of the state
$\rho^{0N}(t)$ is simply given by
$C\left(\rho^{(0N)}(t)\right){=}\left|u_{N1}(t)\right|$.

Several approaches have been put forward in order to maximize this
quantity and in the following we will have a closer look at its
structure.

The transmission amplitude  is given by
\begin{equation}\label{E.uit}
  u_{N1}(t) \equiv \bra{N}\,e^{-i{\cal{H}}t}\ket{1}
  =\sum_{k=1}^N U_{kN}U_{k1}\,e^{-i\omega_k t}~,
\end{equation}
where $U_{ki}$ is the $i$-component of the $k$-eigenvector
entering the  orthogonal matrix that diagonalizes  $\mathbf{T}$,
while $\omega_k$ is the corresponding eigenvalue.

Let us first consider the homogeneous channel, $J_i{=}J{=}1$, and
$h_i{=}h$ $\forall i$ ($J$  will be used as the unit for both
energy and inverse time). In this case,
$U_{ki}{=}\sqrt{\frac{2}{N+1}}\sin \frac{k \pi}{N+1}$ and
$\omega_k{=}\cos\frac{k \pi}{N+1}-h$, $k{=}1,...,N$. Notice first
that an homogeneous magnetic field contributes only with a global
phase to Eq.~(\ref{E.uit}) and, as far as entanglement propagation
is concerned, $h$ can be set hereafter to zero without losing
generality (this holds true  also in the case of non-homogenous
bond constants $J_i$). Because of the non linear functional form
of $\omega_k$ and of the many terms entering the sum in Eq.
(\ref{E.uit}), the entanglement propagation is subject to
dispersion along the chain, although a finite amount of
entanglement  between spin 0 and $N$ is always achievable for any
finite length $N$, {\textit{e.g.}}, for $N{=}10^{3}$ one obtains,
at the optimal time $t^*$, $C\left(\rho_{0N}(t^*)\right){\simeq}
0.135$. The scaling laws of the attainable entanglement and of the
transfer time are given by $C\left(\rho_{0N}(t^*)\right){\sim}1.35
N^{-\frac{1}{3}}$ and $t^*{\sim}N+0.81 N^{\frac{1}{3}}$,
respectively~\cite{Bose2003}.

In the following, we will turn our attention to various proposals
aimed at improving the quality of the transfer. A necessary
condition~\cite{Kay2010} to be fulfilled by the coupling constants
that enter Eq.~(\ref{E.hamilton}) is that the dynamical matrix
$\mathbf{T}$ has to be mirror-symmetric in order to allow
$u_{N1}(t^*)\,{=}1$ at some time $t^*$. This symmetry implies that
the eigenvectors of $\mathbf{T}$  are alternatively symmetric and
antisymmetric with respect to the center of the chain. For
positive (negative) couplings, this yields the condition
$U_{k1}=(-1)^k U_{kN} $, where the eigenvectors are ordered in
such a way that the corresponding eigenvalues are cast in
increasing (decreasing) order. By applying this result to
Eq.~(\ref{E.uit}), we obtain a physically more insightful
expression:
\begin{equation}\label{E.summodes}
u_{N1}(t) {=} \sum_{k=1}^N U_{k1}^2\,e^{i \phi_k(t)} \, ,
\end{equation}
which is a sum over all of the eigen-modes of the channel, each
evolving with its own phase $\phi_k(t){=}\left(k\pi-\omega_k
t\right)$. Because of dispersion, it is quite unlikely that the
phase matching condition $\phi_k(t^*){=}\phi_{k^{'}}(t^*)$, which
would imply perfect state transmission (PST), will be fulfilled
$\forall k,k'$ at some time $t^*$. In order to allow for PST in a
mirror symmetric system, the  state evolution has to be periodic,
which is equivalent to the requirement that the ratios of the
eigenenergy differences have to be rational. It has been shown
that, for linear chains with uniform couplings, this prevents PST
to occur for $N{>}3$ spins~\cite{Christandl2004}. We, then, devote
our attention to spin chains that do not preserve translation
invariance because either the coupling constants or the magnetic
fields are non-uniform.

\subsection{Fully-engineered-coupling scheme}
One possible strategy adopted in order to satisfy the
phase-matching condition is to design the intra-chain couplings in
such a way as to realize a linear spectrum,  $\omega_k{=}\alpha
k$, with $\alpha$ being a constant. By engineering the couplings
according to the rule $J_n{=}\frac{\pi}{N+1}\sqrt{n(N-n)}$,
perfect state transfer can be achieved over arbitrary distances at
time $t^*{=}N{+}1$~\cite{Christandl2004}. Moreover, in
Ref.~\cite{DiFranco2008} it has been shown that the requisite of
the channel initialization can be relaxed if the coupling are
engineered in this way, provided projective measurements are
allowed at the end-chain spins.

The determination of specific coupling schemes of a given
Hamiltonian in order to achieve the desired quantum correlation
transfer process has been addressed also within the Information
Flux approach introduced in Ref.~\cite{DiFranco2007}. It has been
shown that, if the couplings between neighbor spins can be
designed to be non-zero alternatively only along the $X$- or the
$Y$-axes with specific patterns, then PST is found to occur,
together with the generation of maximally entangled states between
the end point of the spin chain itself~\cite{DiFrancoPK2008}.

\subsection{Entanglement transmission as an effective Rabi
oscillation} A completely different strategy to  maximize
$u_{N1}(t)$ is to let only a few eigen-modes enter the sum
in~Eq.(\ref{E.summodes}) with a non-negligible weight. In
particlar, in the ideal case, one would like to have only two
contributing terms \cite{gualdi08}. This can be achieved, for
instance, by coupling the end-point spins (those at sites $1$ and
$N$) very weakly to their neighbors~\cite{Wojcik2005} or by
applying there a strong local magnetic field in the $z$-direction
~\cite{plastina07,FeldmanKZ2010} in an otherwise homogeneous
system. In these cases, only two eigen-modes become really
relevant, corresponding to single excitation states that are
bi-localized around the first and the last spins of the chain,
which take the approximate form
$\ket{\psi_{1,2}}{\simeq}\frac{1}{\sqrt{2}}\left(\ket{1}\pm\ket{N}\right)$.

Rabi-like oscillations, then, take place between these two
eigenstates, and an almost perfect entanglement transfer can be
achieved at the end of the Rabi period, which is inversely
proportional to the difference between the two eigenvalues of the
modes involved, $t^* \sim 1/(\omega_2-\omega_1)$.

\begin{figure}[bt]
\centerline{\psfig{file=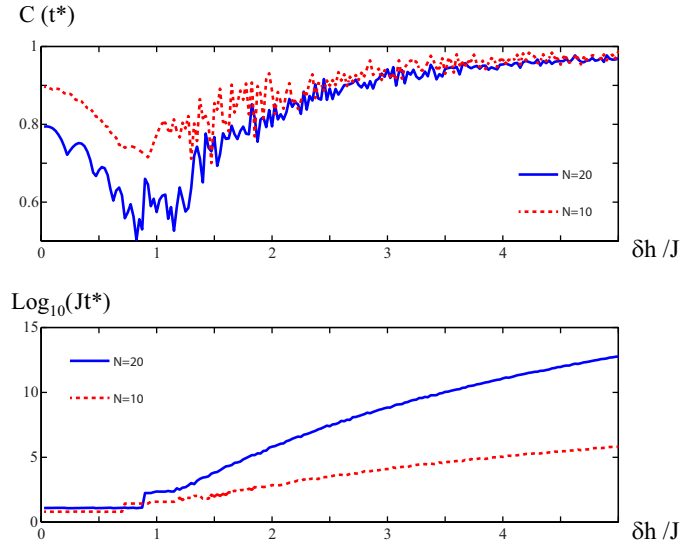,width=0.7\linewidth}}
\vspace*{8pt} \caption{Transferred entanglement (measured by the
concurrence) and optimal transfer times in a spin chain in which
the magnetic field is supplemented by two (equal) local fields
$\delta h$ at the sending and receiving sites. In both panels, the
solid blue line refer to a chain with $N=20$ sites, while the
dashed red one refers to a chain with $N=10$ spins.}
\label{conclogt}
\end{figure}

The approximate expressions given above become exact in the
singular limit in which the couplings become very weak (or the
extra, local magnetic field very strong) with respect to the
exchange constant of the chain. This, however, occurs at the
expense of a much longer transfer time, as in this limit the two
states become quasi-degenerate.

This fact can be seen very clearly in the case in which the
doublet of states that mainly contributes to Eq.
(\ref{E.summodes}) is isolated thanks to an extra local field,
$\delta h$ applied to the first and last spin of the chain. Due to
the presence of this extra field, indeed, two quasi-degenerate
energy levels are detached in energy from the others and acquire
more and more weight in the transition amplitude $u_{N1}$ as
$\delta h$ increases, so that the time evolution that is relevant
for the entanglement transport becomes more and more restricted to
this bi-dimensional subspace. As a result, the transferred
entanglement substantially increases; but this enhancement of the
channel performance comes at the expense of an increased transfer
time. Both of the effects are shown in Fig. (\ref{conclogt}),
where the transferred concurrence is shown, together with the time
it takes for the transfer to occur.

In the case of weak-end-bonds, on the other hand, the exchange
constants connecting spin $1$ and $2$ and spin $N-1$ and $N$, need
to satisfy $J_1{=}J_{N-1}{<<} J / \sqrt N$, so that the transfer
time scales linearly with $N$, see Ref. \cite{Wojcik2005}. This
weak-end-bond approach has also been combined with the one with
engineered coupling strengths, and it has been
shown~\cite{Bruderer2012} that, in this case, the performance of
the transfer protocol has an enhanced resilience to the presence
of random static disorder of the couplings $J^{rnd}_i{=}J_i(1+R)$,
where $R$ is a uniformly distributed random variable.

\subsection{Improving the transmission by optimizing the end-bonds}
The two strategies discussed in the previous sections are in a
certain sense complementary with respect to the aim of maximizing
$u_{N1}(t) $: the first one acts on the phases $\phi_k(t)$  by
fine-tuning all of the couplings of the chain in order to have
appropriate eigenvalues spacing; whereas, the second one acts on
the weights $U_{k1}^2$ by markedly weakening the effective
couplings of spins $1$ and $N$ with the rest of the chain (either
by acting directly on the bond or by introducing an energy
mismatch) in order to obtain only two relevant terms among the
ones that enter the sum in Eq. (\ref{E.summodes}).

The question of whether there exists an optimal tradeoff  between
the  linearization of the spectrum and the density of the excited
modes, with minimal requirements from the engineerization point of
view, has been addressed in Refs.~\cite{BACVV2010,BACVV2011}.
There, it has been shown that the end-point couplingL strengths
influence (and, to some extent, control) at the same time both the
eigenenergy spacing and the width of the excited mode density and
that there exists a finite optimal value $J_1{<}J$ of that
coupling, which allows for a very efficient entanglement transfer.
In particular, the eigenenergies $\omega_{k_n}{=}\cos k_n$
experience a shift (with respect to the values they take with
uniform couplings) towards the center of the spectrum at $k_0 =
\frac{\pi}{2}$ according to the relation
\begin{equation}\label{E.phaseshift}
k_n{=}\frac{n \pi +2 \phi_{k_n}}{N+1},
\end{equation}
where  $\phi_{k_n}{=}k_n-\cot^{-1}\left( \frac{\cot
k_n}{\delta}\right)$ and $n{=}1,...,N$. Because of the lowering of
the coupling, also the  density of the excited modes concentrates
towards the center of the spectrum according to
\begin{equation}\label{E.modedensity}
U_{k1}^2{=}\frac{1}{N+1-2\phi^{'}_{k_n}} \frac{\delta
\left(1+\delta\right)}{\delta^2+\cot^2k_n},
\end{equation}
where $\delta{=}\frac{J_1^2}{2-J_1^2}$ (and $J=1$). The first (and
last) bond coupling strength, thus, can be used as a knob for two
purposes: i) optimizing the spacing of the frequencies
$\omega_{k_n}$, in order to extend the size of the linear zone
around the inflection point $k_0{=}\frac{\pi}{2}$, and ii)
concentrating more excited modes inside this region.

The existence of an optimal value of $J_1$ that maximizes
$u_{N1}(t)$ is due to the interplay between two conflicting
effects: the ``linearization'' effect, that becomes less
pronounced as $J_1$ is lowered, and the ``concentration'' effect
that, on the contrary, is found to occur more and more as $J_1$ is
lowered.

Once the end-point couplings are set to their optimal values, an
almost coherent dynamics arises, yielding an high quality transfer
of entanglement in a time $t^*{\simeq}N{+}1.89 N^{\frac{1}{3}}$,
where the increased delay with respect to the perfectly ballistic
transfer time $N$ is precisely due to the fact that the first and
last ``steps'' take more time as $J_1{<}1$. The transferred
entanglement can achieve very high values, {\textit{e.g.}},
$C\left(\rho_{0N}(t^*)\right){\simeq} 0.89$ for $N{=}10^3$.
Remarkably, also in the limit $N{\rightarrow}\infty$, entanglement
transfer takes places with an efficiency around 85$\%$ and the
optimal coupling scales as $J_1^{opt}{\simeq}N^{-\frac{1}{6}}$,
whereas for the all-uniform-coupling case,
$C\left(\rho_{0N}(t^*)\right){\rightarrow}0$.

\begin{figure}[bt]
\centerline{\psfig{file=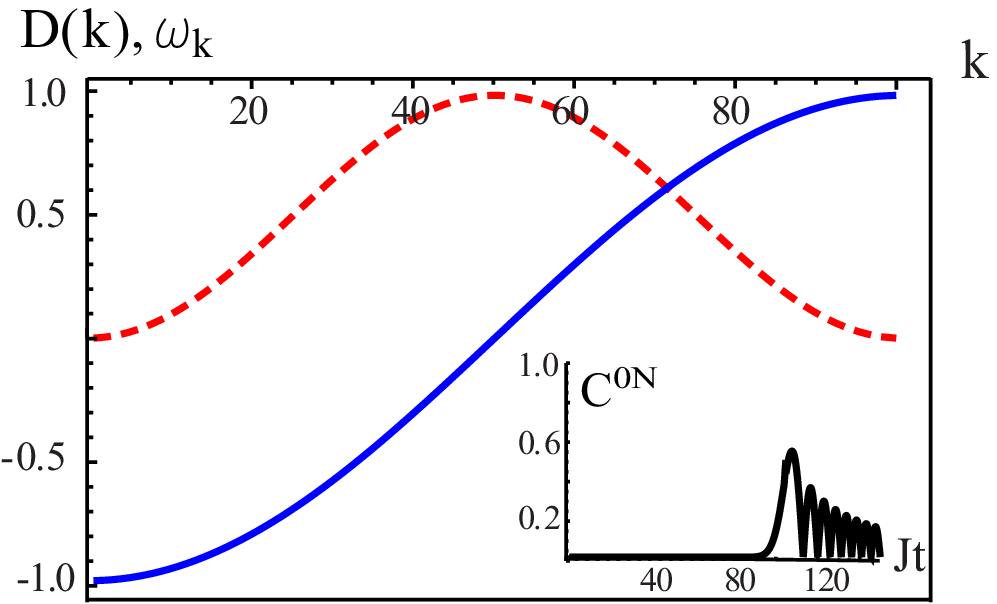,width=0.38\linewidth}
\psfig{file=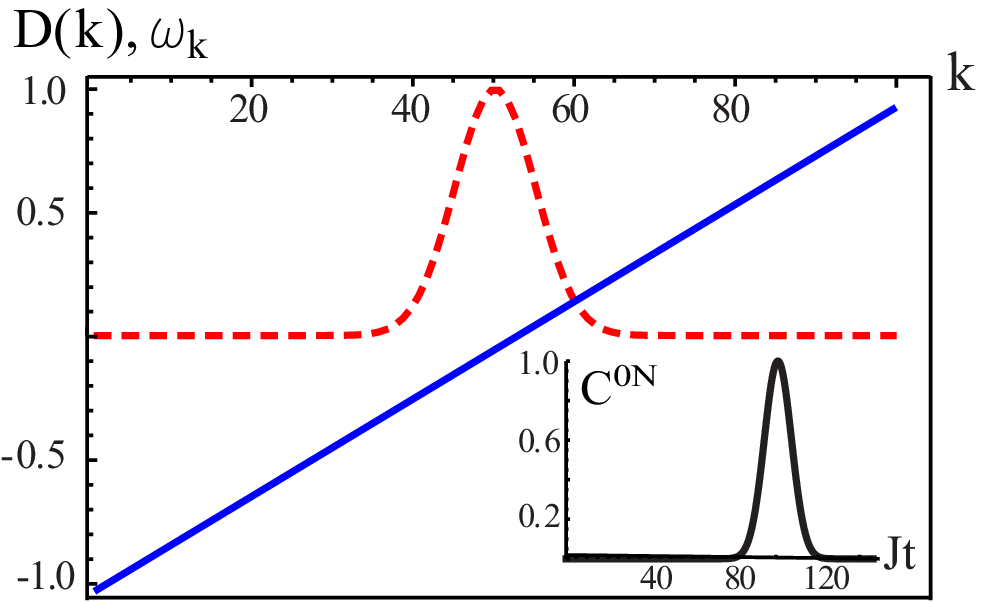,width=0.38\linewidth}} \vspace*{8pt}
\centerline{\psfig{file=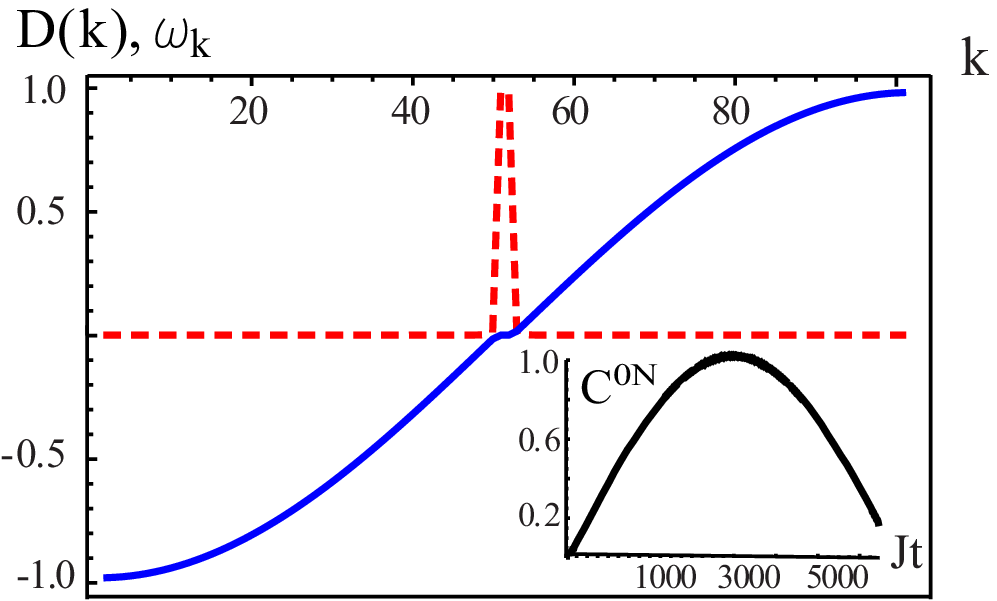,width=0.38\linewidth}
\psfig{file=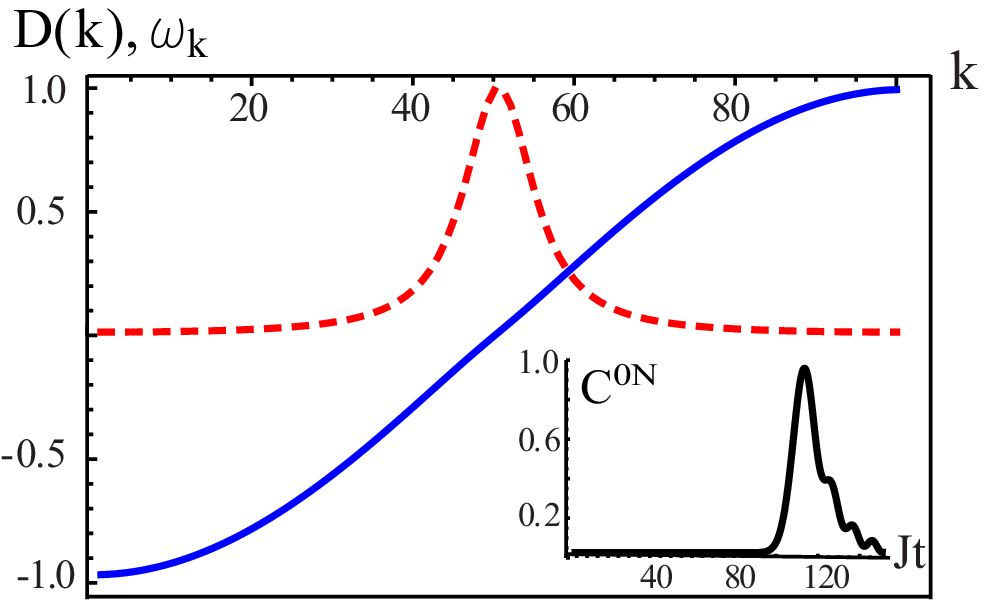,width=0.38\linewidth}} \vspace*{8pt}
\caption{First row:  normalized density of excited modes $
D(k){\equiv}\frac{U_{k1}^2}{\max_k\left[U_{k1}^2\right]}$(red
dashed line) and  eigenvalues  $\omega_k$, $k{=}1,...,N$ (blue
solid line) in the case of the uniform (left plot), and the fully
engineered chains (right plot). Second row: same quantities for
the weak-end-bond case, with $J_1{=}10^{-2}$ (left plot) and for
optimal couplings $J_1{=}0.49$ (right plot). The insets show the
time evolution of the concurrence between spin $0$ and $N$ in the
corresponding regimes.} \label{F.denspe}
\end{figure}

These considerations are summarized in Fig.~\ref{F.denspe}, where
the (normalized) density of excited modes ${
D}(k){\equiv}\frac{U_{k1}^2}{\max_k\left[U_{k1}^2\right]}$ and the
eigenenergies $\omega_k$ entering Eq.~(\ref{E.summodes}) are
reported for a chain of $N{=}100$ spins. The case of uniform
couplings is compared with the one in which the couplings are
fully optimized, and with that in which only weak end-bonds are
included. It is evident that with uniform couplings, the
distribution of modes is quite broad and the dispersion relation
is manifestly non linear, which yields a poor quality entanglement
transfer across long chains. On the contrary, in a chain with
fully engineered couplings, the single particle spectrum is
completely linear and the density of mode is gaussian, which
guarantees a perfect transfer. In the case of a chain with weak
end-couplings, there are two quasi-degenerate dominant modes
entering the dynamics, which implies that a good transfer quality
requires very long times. Finally, the optimal coupling scheme
with (slightly) weaker end bonds accomplishes both a shrinking of
the relevant modes and an approximate linearization of their
dispersion relation

The effect of static disorder on the couplings within the transfer
channel, has been also taken into account. The presence of a
random contribution in the exchange constants,
$J^{rnd}_i{=}J_i+\delta J_i$, $i{=}2,...,N-2$, has been
extensively investigated in~\cite{ZwickASO2011,ZwickASO2012} for
different kinds of static disorder $\delta J_i$ where a detailed
comparison of the transfer performances of the fully-engineered
chain, of the chain with weak-ends and of that with optimal
couplings is presented.

A further relevant improvement of the optimal coupling transfer
scheme has been put forward in Ref.~\cite{ApollaroBCVV2012} by
allowing also the second- (and last-but-one) coupling $J_2$ (equal
to $J_{N-2}$) to be set at an optimal value, with $J_1{<}J_2{<}J$.
With this additional optimization, the two competing effects of
the deformation of the eigenvalue spacing and of the shrinking of
the mode distribution can be handled independently, instead of
being both controlled by $J_1$ as in the original scheme. It turns
out that the eigenvalues are given by the same expression of
Eq.~(\ref{E.phaseshift}), but with the phase shifts $\phi_{k_n}$
that are mainly ruled by $J_2$, whereas the mode density
$U_{k1}^2$ is mainly controlled by $J_1$. As a result, a
considerably higher amount of entanglement can be transferred
across very long chains, in a time that scales as
$t^*{\simeq}N{+}3.24N^{\frac{1}{3}}$. In the limit
$N{\rightarrow}\infty$, entanglement transfer takes places with an
efficiency of more than 99$\%$ provided that $J_1$ and $J_2$ are
set to their optimal values, which scale with the size of the
chain as $J_1^{opt}{\simeq}N^{-\frac{1}{3}}$ and
$J_2^{opt}{\simeq}N^{-\frac{1}{6}}$.

\subsection{Propagation of quantum discord}
While most of the work in the context of spin-channels  has
focused on the study of the propagation of entanglement in such
media, it is now widely accepted that nonclassical correlations do
not reduce to just quantum entanglement. Quantum discord
\cite{disco} and measurement-induced disturbance \cite{mid}, just
to mention two well known examples, are able to quantify the
nonclassical correlation content of a given quantum state well
beyond entanglement. Although their role in quantum information
processing has not been fully clarified yet, the interest in their
properties has constantly increased in the last few years as they
are understood as general indicators of the quantumness of a
state.

In Ref.~\cite{Campbell2011}, the authors have addressed the
question of whether or not the fundamentally conceptual difference
between entanglement ${\cal E}$ and discord ${\cal D}$ leaves
signatures in the way such non-classical quantities are
transferred across a spin channel, with the same setting described
above, Fig. (\ref{F.qchan}). The question makes full sense  both
for a pure and a mixed initial state $\rho^{(01)}$, as in both
cases the output state of the transfer protocol needs not be pure,
so that, in general, ${\cal D}\ne {\cal E}$. To analyze the
transfer efficiency, the best choice is to use as figure of merit
the re-scaled quantities $\tilde{\cal D}\,{=}\,{\cal D}/{\cal
D}_{(01)}$ and $\tilde{\cal E}\,{=}\,{\cal E}/{\cal E}_{(01)}$,
where ${\cal D}$ is the discord contained in the state
$\rho^{(0N)}(t^*)$, while ${\cal E}$ is the entanglement of
formation of the same state (which is chosen here in place of the
concurrence in order to have two entropic-like quantities to
compare), and their values are renormalized with respect to the
amount of the corresponding quantity in the initial state.

\begin{figure}[bt]
\centerline{\psfig{file=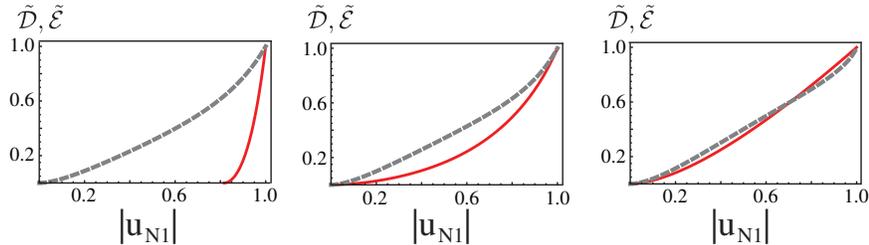,width=0.9\linewidth}}
\vspace*{8pt} \caption{Comparison between the re-scaled discord
$\tilde{\cal D}$ (grey dashed line) and entanglement of formation
$\tilde{\cal E}$ (red solid one) propagated across a chain of
given transition amplitude $|u_{N1}|$. The spin-pair $(1,0)$ is
prepared in a Werner state with $a\,{=}\,0.4,0.7$ and $1$ [panel
{\bf (a)}, {\bf (b)} and {\bf (c)}, respectively].}
\label{figureW}
\end{figure}

As possible input states of the qubit  pair~$(0,1)$, pure and
$X$-type mixed states of the form reported in Eq.~(\ref{E.Xrho})
have been considered, the latter being of particular relevance
because both maximally entangled mixed states and maximally
discorded mixed states fall into this class~\cite{Mdms,MEMS}.
Despite the substantial differences among the various input
states, a common feature of the transport process is that the
discord propagates better than entanglement for a wide range of
input states and working conditions of the channel; and that
$\tilde{\cal E}$ exceeds $\tilde{\cal D}$ only when both the
transition amplitude and the purity of the initial state are large
enough. A representative example of this behavior can be given by
choosing $\rho^{(01)}(0)$ to be a Werner state~\cite{werner},
whose non-null matrix elements in Eq.~(\ref{E.Xrho}) are given by
$\rho_{11}{=}\rho_{44}{=}\frac{1+a}{4},
\rho_{22}{=}\rho_{33}{=}\frac{1-a}{4}, \rho_{14}{=}\frac{a}{2}$,
where $-1/3\,{\le}\,{a}\,{\le}\,{1}$ and which is entangled for
$a{\ge}1/3$. As shown in Fig.~(\ref{figureW}), for values mildly
larger than this threshold, where the purity of the state is small
and so is its entanglement, very large values of $|u_{N1}|$ are
required in order to actually transport entanglement. On the other
hand, the discord ${\cal D}$ is non-null for any value of
$|u_{N1}|$ and irrespectively of the initial discord content of
the pair $(0,1)$. The relative discrepancy between the two figures
of merit is in general very large and decreases only for almost
ideal transport conditions. As $a\rightarrow1$, that is, by
increasing the purity of the state, more entanglement is present
at the beginning and larger and larger fractions of it are
transported, even for small transition amplitudes $|u_{N1}|$.
Thus, for increasingly pure initial states, the differences
between the two non-classicality indicators are more and more
reduced for a good channel. In the limit of $a{=}1$, which makes
$\rho^{(01)}$ a maximally entangled pure state, discord is
overtaken by the entanglement of formation at $|u|\,{\ge}\,1/\sqrt
2$.

Moreover, for the case of a spin chain with uniform couplings, we
have seen that the entanglement of formation of $\rho^{(0N)}(t)$
is invariant with respect to the presence of an homogeneous
magnetic field; whereas, for the quantum discord, this is not the
case if the initial state $\rho^{(01)}(0)$ displays different
values of the spin-spin correlations on the $xy$-plane; that is,
if
$\exval{\sigma_0^x\sigma_1^x}{\neq}\exval{\sigma_0^y\sigma_1^y}$.
In fact, the magnetic field increases the amount of discord that
can be obtained between $0$ and $N$ as compared to the case with
$h{=}0$.

The formal structure of Eq.~(\ref{E.krauss}) can be employed also
for the study of the storage of the quantum correlations shared by
spin 0 and 1 in the presence of an environment acting only on
qubit 1. Thus, the same description presented up to now can be
adopted to discuss the preservation of entanglement and of quantum
discord in the case in which one of the initially correlated
members is acted upon by an environment, constituted in our case
by the spins at sites $2, \ldots , N$. Since entanglement cannot
increase by LOCC, at variance with Discord, strikingly different
dynamical behaviors of these two kind of quantum correlations take
place in open systems. In fact in
Refs.~\cite{Ciccarello2012,Streltsov2011} it is shown that a local
and memoryless environment can indeed generate quantum discord
starting from a purely classical state. Moreover, under the effect
of Markovian environments, the two quantities can display very
different behaviors as entanglement can experience sudden death,
while quantum discord can vanish only
asymptotically~\cite{Werlang2009} or even stay constant for a
finite time interval~\cite{Mazzola2010}. These different dynamical
behaviors are reproduced in the case of the effective amplitude
damping channel obtained by considering our spin chain
\cite{Campbell2011}.

\section{Anisotropic models}
\label{animodel} In the protocols described so far, the initial
state of the quantum channel,  (that is, the state of the spins
residing at sites ranging from  $2$ to $N$) has been chosen to be
initialized in the zero-particle sector. As a result, the presence
of an additional spin-spin interaction along the $z$-direction in
the Hamiltonian $\hat{\cal H}$, that is introduced in Eq.
(\ref{E.hamilton}) by letting $\Delta{\neq}0$, does not affect
significantly the transfer process. This is a consequence of the
fact that the dynamical effect of the additional coupling term is
equivalent to that of an overall uniform magnetic field, applied
to every spin but for the ones at end-points; which implies that
the dynamics still remains restricted to the zero- and
single-excitation sectors of the total Hilbert space.

\begin{figure}[b]
\centerline{\psfig{file=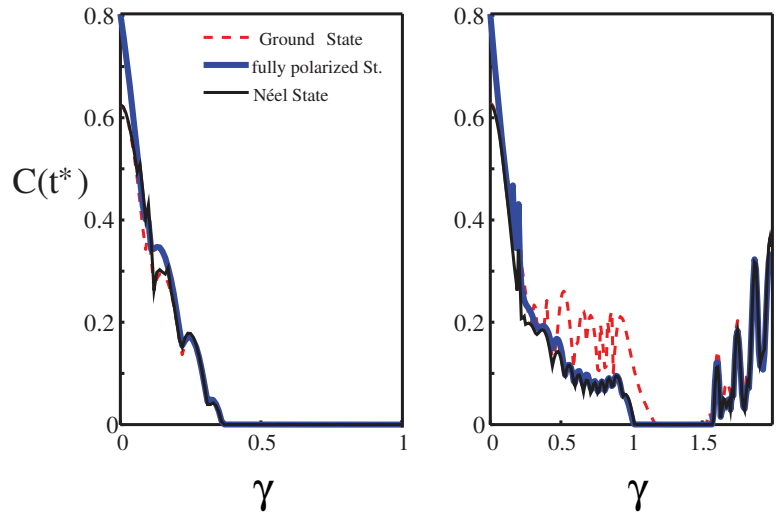,width=0.8\linewidth}}
\caption{Concurrence vs  in-plane anisotropy $\gamma$, for $h{=}0$
(left) and $h{=}0.5$ (right) in a chain of length $N{=}20$ for
different initial states: ground state (red-dashed line),
ferromagnetic state (blue, thick line) and N\'{e}el state (black,
thin line). }\label{F.pra83_062328_f4}
\end{figure}

On the other hand, if one considers a spin chain that is not fully
polarized at the beginning, but that, rather, is prepared, e.g.,
in its ground state, then the transfer of quantum correlations
will strongly depend on the value of $\Delta$ \cite{Bayat2010}. In
particular, the best working point (as far as both the amount of
the transferred entanglement and the transfer speed are concerned)
turns out to be the $\Delta{=}1$ anti-ferromagnetic point. An
interesting question is whether different channel initializations
can result in an improvement of the transfer efficiency. This
topic has been addressed in Ref.~\cite{Bayat2011}, where a
comprehensive analysis of the transport of entanglement has been
performed for the anti-ferromagnetic ($J>0$) $XY$ model (with
$\gamma{\neq} 0$, $\Delta{=}0$) and for the  $XXZ$ model
($\gamma{=} 0$, $\Delta{\neq}0$), by considering different
channel's initial states: the ferromagnetic polarized state, in
which all of the spins are parallel to each other; the N\'{e}el
state, in which neighboring spins are antiparallel; and the ground
state. As shown in Fig.~\ref{F.pra83_062328_f4} for a chain made
up of $N=20$ spins, the entanglement turns out to be better
transmitted in the $XX$ case ($\gamma{=}0$), whereas, in the $XY$
model, the transfer capability of the model falls to zero already
for moderate values of the anisotropy $\gamma$. Furthermore, for
the $XX$ model, among the considered initial states, the
ferromagnetic one is found to be the most efficient. Nevertheless,
the presence of a uniform magnetic field $h$ in the $Z$-direction
enables to obtain a finite amount of entanglement between spin $0$
and $N$ also in the presence of strong anisotropy and it turns out
that the ground state becomes the most efficient initial state
when the field is present.

\begin{figure}[b]
\centerline{\psfig{file=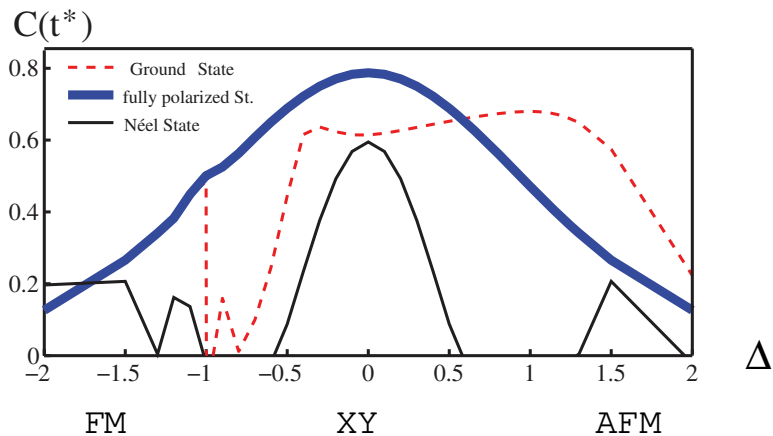,width=0.8\linewidth}}
\caption{Attainable Concurrence vs $\Delta$ over the complete
phase diagram of the $XXZ$ model for different initial states. }
\label{F.pra83_062328_f7b}
\end{figure}

In the case of the $XXZ$ channel, which, unlike the models with
$\Delta{=}0$, doesn't map into a free fermion system, one has to
take into account also the scattering effects between the
interacting excitations which gives rise to a more complex
behavior of the transmission efficiency, which depends on the
phase of the model: different initial states exhibit a strong
dependence of the transfer quality for different $\Delta$'s, as
reported in Fig.~\ref{F.pra83_062328_f7b} (see Ref.
\cite{Bayat2011} for more details).

\section{Concluding remarks}
\label{conclurem}  From the point of view of coherent information
processing, the nontrivial dispersion properties of spin networks
represent an interesting opportunity for their use as
short-distance communication channels for the interconnections
among on-chip nodes in the next generation of information
processing devices. In this context, in this paper we have
reviewed various properties of finite one dimensional spin
systems, employable as quantum channels to transmit quantum
correlations (and, in particular, entanglement) from one end to
the other. The efficiency with which such correlations are
propagated across the spin chain appears to be larger for a system
with $XX$ interactions between spins initially prepared in the
fully polarized state. In such a case, the transfer properties
only depend on the modulus of the transfer function $u_{N1}(t)$,
given in Eq. (\ref{E.summodes}), that describes the propagation of
a localized single-particle (fermion) excitation, and whose value
can be adjusted to some extent by employing the coupling constants
between neighboring spins as control parameters. In particular, a
high quality transmission in an almost ballistic time can be
obtained with minimal requirements in terms of the degree of
engineering of the chain, if one adopts the optimized end-bond
coupling scheme, in which the first and the last bonds are
weakened (with respect to all of the others) in such a way as to
approximately linearize the  excitation spectrum of the
 modes that effectively contribute to the propagation.

To finish this summary, we just mention that the models that have
been considered can be experimentally realized in various ways,
ranging from NMR samples to atoms loaded into one-dimensional
optical lattices that are able to simulate spin Hamiltonians with
a very high degree of accuracy \cite{bloch}.

\section{Acknowledgement}
TJGA is supported by the European Commission, the European Social
Fund and the Region Calabria through the program POR Calabria FSE
2007-2013-Asse IV Capitale Umano-Obiettivo Operativo M2.

\end{document}